\documentclass[12pt]{amsart}
\usepackage{amssymb}
\usepackage{color}
\usepackage{amsmath,epic,curves}
\usepackage[english]{babel}
\usepackage{graphicx}

\newtheorem{claim}{}[section]
\newtheorem{theorem}[claim]{Theorem}

\begin{document}
\baselineskip 6.0 truemm
\parindent 1.5 true pc

\newcommand\lan{\langle}
\newcommand\ran{\rangle}
\newcommand\tr{{\text{\rm Tr}}\,}
\newcommand\ot{\otimes}
\newcommand\ol{\overline}
\newcommand\join{\vee}
\newcommand\meet{\wedge}
\renewcommand\ker{{\text{\rm Ker}}\,}
\newcommand\im{{\text{\rm Im}}\,}
\newcommand\id{{\text{\rm id}}}
\newcommand\tp{{\text{\rm tp}}}
\newcommand\pr{\prime}
\newcommand\e{\epsilon}
\newcommand\la{\lambda}
\newcommand\inte{{\text{\rm int}}\,}
\newcommand\ttt{{\text{\rm t}}}
\newcommand\spa{{\text{\rm span}}\,}
\newcommand\conv{{\text{\rm conv}}\,}
\newcommand\rank{\ {\text{\rm rank of}}\ }
\newcommand\re{{\text{\rm Re}}\,}
\newcommand\ppt{\mathbb T}
\newcommand\rk{{\text{\rm rank}}\,}
\newcommand\zzzz{\par\bigskip\texttt{... to be filled up}....\bigskip\par}

\title{Entanglement witnesses arising from Choi type positive linear maps}

\author{Kil-Chan Ha}
\address{Faculty of Mathematics and Statistics, Sejong University, Seoul 143-747, Korea}

\author{Seung-Hyeok Kye}
\address{Department of Mathematics and Institute of Mathematics\\Seoul National University\\Seoul 151-742, Korea}

\thanks{KCH is partially supported by NRFK 2012-0002600. SHK is partially supported by NRFK 2012-0000939}



\begin{abstract}
We construct optimal PPTES witnesses to detect $3\otimes 3$ PPT
entangled edge states of type $(6,8)$ constructed recently \cite{kye_osaka}. To do
this, we consider positive linear maps which are variants of the
Choi type map involving complex numbers, and examine several notions
related to optimality for those entanglement witnesses.
Through the discussion, we suggest a method to check the optimality
of entanglement witnesses without the spanning property.
\end{abstract}

\maketitle

\section{Introduction}

The notion of quantum entanglement plays a key role in the current
study of quantum information and quantum computation theory. There
are two main criteria to distinguish entanglement from separable
states: The PPT criterion \cite{choi-ppt, peres} tells us that the
partial transpose of a separable state is positive, that is,
positive semi-definite. The converse is not true in general by a
work of Woronowicz \cite{woronowicz} who gave an example of a
$2\otimes 4$ PPT entangled state. Such examples were also given in
\cite{choi-ppt,stormer82} for the $3\otimes 3$ cases, in the early
eighties. Another complete criterion was given by Horodecki's
\cite{horo-1} using positive linear maps between matrix algebras,
and this was formulated as the notion of entanglement witnesses
\cite{terhal}. This is equivalent to the duality theory
\cite{eom-kye} between positivity of linear maps and separability of
block matrices, through the Jamio\l kowski-Choi isomorphism
\cite{choi75-10,jami}. Through this isomorphism, an entanglement witness is just a
positive linear map which is not completely positive. We refer to
\cite{ssz,ZB} for systematic approaches to the duality using the JC
isomorphism.

For a linear map $\phi$ from the $C^*$-algebra $M_m$ of all $m\times m$ matrices into $M_n$, the Choi matrix $C_\phi$ of $\phi$
is given by
$$
C_\phi=\sum_{i,j}^n e_{ij}\ot \phi(e_{ij})\in M_m\ot M_n,
$$
where $e_{ij}=| i\ran\lan j| $ is the usual matrix units in $M_m$. The correspondence $\phi\mapsto C_\phi$ is called the JC isomorphism.
It is known that $\phi$ is positive if and only if $C_\phi$ is block-positive, and $\phi$ is completely positive if and only if
$C_\phi$ is positive.
For a linear map $\phi:M_m\to M_n$ and a block matrix $A\in M_m\otimes M_n$, we define the bilinear pairing by
$$
\lan A,\phi\ran=\tr(AC_\phi^\ttt).
$$
It turns out that $A$ is separable if and only if $\lan A,\phi\ran \ge 0$ for every positive map $\phi$, and
$A$ is of PPT if and only if $\lan A,\phi\ran \ge 0$ for every decomposable positive map $\phi$.
Therefore, every entangled state $A$ is detected by a positive linear map $\phi$ in the sense that
$\lan A,\phi\ran<0$, and every PPT entangled state is detected by an indecomposable positive linear map.
A positive linear map is said to be an optimal entanglement witness if it detects a maximal set of entanglement,
and an optimal PPTES witness if it detects a maximal set of PPT entanglement, as was introduced in \cite{lew00}.
See also \cite{ha_kye_opt_ind} for the terminology.
For a given entangled state, it is easy to find an entanglement witness to detect it, as it was suggested in \cite{lew00}.
But, it is not clear at all how to construct an optimal entanglement witness to detect a given entangled state.
The primary purpose of this note is to construct optimal PPTES witnesses which detect PPT entangled edge states constructed in
\cite{kye_osaka}. These states are the first examples of two qutrit PPT entangled edge states of type $(6,8)$, whose existence
had been a long standing question \cite{sbl}. We also suggest a method to check the optimality of entanglement witness
without the spanning property.

For nonnegative real numbers $a,b,c$ and $-\pi\le\theta\le\pi$,
we consider the map $\Phi[a,b,c;\theta]$ between $M_3$ defined by
$$
\Phi[a,b,c;\theta](X)=\\
\begin{pmatrix}
ax_{11}+bx_{22}+cx_{33} & -e^{i\theta}x_{12} & -e^{-i\theta}x_{13} \\
-e^{-i\theta}x_{21} & cx_{11}+ax_{22}+bx_{33} & -e^{i\theta}x_{23} \\
-e^{i\theta}x_{31} & -e^{-i\theta}x_{32} & bx_{11}+cx_{22}+ax_{33}
\end{pmatrix},
$$
for $X\in M_3$. Note that the Choi matrix $C_\Phi$ of the map $\Phi[a,b,c;\theta]$ is given by
$$
W[a,b,c;\theta]=\left(
\begin{array}{ccccccccccc}
a     &\cdot   &\cdot  &\cdot  &-e^{i\theta}     &\cdot   &\cdot   &\cdot  &-e^{-i\theta}     \\
\cdot   &c &\cdot    &\cdot    &\cdot   &\cdot &\cdot &\cdot     &\cdot   \\
\cdot  &\cdot    &b &\cdot &\cdot  &\cdot    &\cdot    &\cdot &\cdot  \\
\cdot  &\cdot    &\cdot &b &\cdot  &\cdot    &\cdot    &\cdot &\cdot  \\
-e^{-i\theta}     &\cdot   &\cdot  &\cdot  &a     &\cdot   &\cdot   &\cdot  &-e^{i\theta}     \\
\cdot   &\cdot &\cdot    &\cdot    &\cdot   &c &\cdot &\cdot    &\cdot   \\
\cdot   &\cdot &\cdot    &\cdot    &\cdot   &\cdot &c &\cdot    &\cdot   \\
\cdot  &\cdot    &\cdot &\cdot &\cdot  &\cdot    &\cdot    &b &\cdot  \\
-e^{i\theta}     &\cdot   &\cdot  &\cdot  &-e^{-i\theta}     &\cdot   &\cdot   &\cdot  &a
\end{array}
\right).
$$
The map of the form $\Phi[a,b,c;0]$ and its variants have been investigated by many authors in various contexts,
as it was summarized in \cite{ha_kye_opt_ind}.
We just note here that $W[a,b,c;0]$ is separable if and only if it is of PPT \cite{kye_osaka}.
On the other hand, many interesting examples of PPT states are of the form $W[a,b,c;\theta]$. For example,
the PPT entangled states \cite{stormer82} considered in the early eighties are just $W[1,b,\frac 1b,\pi]$
which turn out to be  PPT entangled edge states of type $(6,7)$.
These states were reconstructed systematically from indecomposable positive linear maps
together with other types of PPT entangled edge states
\cite{ha-kye-2}. The PPT entangled edge states of type $(6,8)$ constructed in \cite{kye_osaka}
are given by $W[e^{i\theta}+e^{-i\theta},b,\frac 1b;\theta]$
for $-\frac\pi 3<\theta<\frac\pi 3$ and $\theta\neq 0$. Recently, the authors \cite{ha_kye_geom} analyzed $W[a,b,c;\pi]$
to understand the boundary structures between separability and inseparability among PPT states.

For indecomposable positive linear maps, we have to be very careful to use the term \lq optimal\rq\
as was noticed in \cite{ha_kye_opt_ind}.
We denote by $\mathbb P_1$ the convex cone of all positive linear maps.
Recall that a positive map $\phi$ is an optimal (respectively co-optimal) entanglement witness
if and only if the smallest face of $\mathbb P_1$ determined
by $\phi$ has no completely positive (respectively completely copositive) map.
We also note that $\phi$ has the spanning property (respectively co-spanning property)
if and only if the smallest exposed face of $\mathbb P_1$ determined by $\phi$
has no completely positive (respectively completely copositive) map.
We say that $\phi$ is bi-optimal  if it is both optimal and co-optimal. The term bi-spanning is defined similarly.

After we give conditions on parameters $a,b,c$ and $\theta$ for which the map $\Phi[a,b,c;\theta]$
is a positive linear map in the next section, we characterize for each fixed $\theta$ the facial structures of the
$3$-dimensional convex body
representing the positivity  in Section 3. From this facial structures, it is clear that some of
positive maps are not optimal, and/or not co-optimal.
We examine the spanning and co-spanning properties
for the map $\Phi[a,b,c;\theta]$ in Section 4, and check various notions of optimality for all cases in Section 5.
To do this, we suggest a more efficient method to check the optimality of an entanglement witness
when it has not the spanning property.
In the Section 6, we find optimal entanglement witnesses to detects PPT entangled edge states \cite{kye_osaka} of type $(6,8)$.
We conclude this note to report that our constructions give counter-examples to the SPA conjecture \cite{korbicz}.

\section{Positivity}

To begin with, we first find the conditions for complete positivity and complete copositivity.
We note that the map $\Phi[a,b,c;\theta]$ is completely positive if and only if
$W[a,b,c;\theta]$ is positive if and only if
the following $3\times 3$ matrix
$$
P[a,\theta]=
\left(
\begin{matrix}
a & -e^{i\theta} & -e^{-i\theta}\\
-e^{-i\theta} & a & -e^{i\theta}\\
-e^{i\theta} & -e^{-i\theta} & a
\end{matrix}
\right)
$$
is positive. We mention again that \lq positivity\rq\ of matrices means the positive semi-definiteness,
throughout this note.
We see that the polynomial
$$
\begin{aligned}
\det P[a,\theta]
&=a^3-3a-(e^{3i\theta}+e^{-3i\theta})\\
&=[a-(e^{i\theta}+e^{-i\theta})]\,[(a^2+a(e^{i\theta}+e^{-i\theta})+(e^{2i\theta}+e^{-2i\theta}-1)]
\end{aligned}
$$
has the following three real zeroes:
$$
q_{(\theta-\frac 23 \pi)}=e^{i(\theta-\frac 23 \pi)}+e^{-i(\theta-\frac23 \pi)},\quad
q_\theta=e^{i\theta}+e^{-i\theta},\quad
q_{(\theta+\frac 23 \pi)}=e^{i(\theta+\frac 23 \pi)}+e^{-i(\theta+\frac 23 \pi)}.
$$
We denote by
$$
p_\theta=\max\{ q_{(\theta-\frac 23 \pi)}, q_\theta, q_{(\theta+\frac 23 \pi)}\}.
$$
\begin{figure}[h!]
\begin{center}
\includegraphics[scale=0.7]{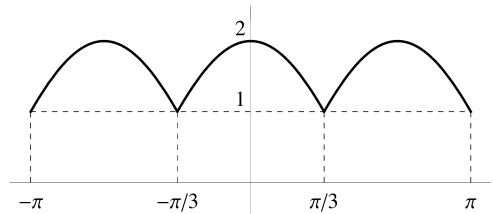}
\end{center}
\caption{The graph of $y=p_{\theta}$. }
\end{figure}

Then we see that $1\le p_\theta\le 2$ for each $\theta$, and
$P[a,\theta]$ is positive if and only if
\begin{equation}\label{cp}
a\ge p_\theta,
\end{equation}
if and only if $\Phi[a,b,c;\theta]$  is completely positive. We note that $p_\theta=2$ if and only if
$\theta=0,\pm\frac {2}3\pi$, and $p_\theta=1$ if and only if $\theta=\pm \frac \pi 3,\pm\pi$.
It is easy to see that
the map $\Phi[a,b,c;\theta]$  is completely copositive if and only if
\begin{equation}\label{ccp}
bc\ge 1.
\end{equation}
\begin{theorem}
The map $\Phi[a,b,c;\theta]$ is completely positive if and only if the condition {\rm (\ref{cp})} holds,
and completely copositive if and only if the condition {\rm (\ref{ccp})} holds.
\end{theorem}

In order to get a necessary condition for the positivity of $\Phi[a,b,c;\theta]$, we note that
a linear map $\phi$ is positive if and only if
$\lan zz^*,\phi\ran\ge 0$ for every product vector $z=x\otimes y\in\mathbb C^m\ot\mathbb C^n$.
Here, $z$ is considered as a column vector, and so $zz^*$ belongs to $M_m\otimes M_n$.
We write
$$
z=x\otimes y=(
x_1y_1, x_1y_2, x_1y_3\,;\,
x_2y_1, x_2y_2, x_2y_3\,;\,
x_3y_1, x_3y_2, x_3y_3)^{\rm t}.
$$
By a direct calculation, we see that the pairing $\lan zz^*, \Phi[a,b,c;\theta]\ran$  is equal to
$$
\begin{aligned}
&a(|x_1y_1|^2+|x_2y_2|^2+|x_3y_3|^2)
+
b(|x_1y_3|^2+|x_2y_1|^2+|x_3y_2|^2)
+
c(|x_1y_2|^2+|x_2y_3|^2+|x_3y_1|^2)\\
&-e^{i\theta} x_1 y_1 \bar x_2 \bar y_2  -e^{-i\theta} \bar x_1 \bar y_1 x_2 y_2
-e^{i\theta} x_2 y_2 \bar x_3 \bar y_3  -e^{-i\theta} \bar x_2 \bar y_2 x_3 y_3
-e^{i\theta} x_3 y_3 \bar x_1 \bar y_1  -e^{-i\theta} \bar x_3 \bar y_3 x_1 y_1.\\
\end{aligned}
$$
From now on, we suppose that  $\Phi[a,b,c;\theta]$ is positive, and put the product vectors
$$
(1,1,1)^{\rm t}\ot(1,1,1)^{\rm t},\quad
(e^{\frac 23\pi i},1,1)^{\rm t}\ot(1,1,1)^{\rm t},\quad
(e^{-\frac 23\pi i},1,1)^{\rm t}\ot(1,1,1)^{\rm t}
$$
in the above quantity, to get the following necessary condition
\begin{equation}\label{p1}
a+b+c\ge p_\theta.
\end{equation}
We also take product vectors
\[
(\sqrt te^{-i\theta},t,0)^{\rm t}\ot (\sqrt t,1,0)^{\rm t},\quad
(\sqrt te^{-i\theta},t,0)^{\rm t}\ot (\sqrt t,e^{\frac 23\pi i},0)^{\rm t},\quad
(\sqrt te^{-i\theta},t,0)^{\rm t}\ot (\sqrt t,e^{-\frac 23\pi i},0)^{\rm t}
\]
for $t\ge 0$, to get the condition $2at^2+ct+bt^3\ge  2t^2$ for each
$t\ge 0$
if and only if
\begin{equation}\label{p2}
a\le 1 \Longrightarrow\
bc\ge (1-a)^2.
\end{equation}
Therefore, we get necessary conditions (\ref{p1}) and (\ref{p2}) for the positivity of  $\Phi[a,b,c;\theta]$.

Now, we proceed to show that two conditions (\ref{p1}) and (\ref{p2}) are sufficient for positivity.
Note that $\Phi[a,b,c;\theta]$ is positive if and only if the matrix
\begin{equation}\label{mat}
\begin{pmatrix}
a|x|^2+b|y|^2+c|z|^2 & -e^{i\theta}x\bar y & -e^{-i\theta}x\bar z \\
-e^{-i\theta}y\bar x & c|x|^2+a|y|^2+b|z|^2 & -e^{i\theta}y\bar z \\
-e^{i\theta}z\bar x & -e^{-i\theta}z\bar y & b|x|^2+c|y|^2+a|z|^2
\end{pmatrix}
\end{equation}
is positive semi-definite for any $(x,y,z)\in\mathbb C^3$.
We first consider the determinant
$$
\begin{aligned}
(a|x|^2+b|y|^2+c|z|^2)(c|x|^2+a|y|^2+b|z|^2)(b|x|^2+c|y|^2+a|z|^2)
-(e^{3i\theta}+e^{-3i\theta})|xyz|^2\\
-(a|x|^2+b|y|^2+c|z|^2)|yz|^2-(c|x|^2+a|y|^2+b|z|^2)|zx|^2
-(b|x|^2+c|y|^2+a|z|^2)|xy|^2.
\end{aligned}
$$
We may replace $|x|^2, |y|^2$ and $|z|^2$ by nonnegative $x,y$ and $z$ to get
$$
\begin{aligned}
F(x,y,z):=(ax+by+cz)(cx+ay+bz)(bx+cy+az)
-(e^{3i\theta}+e^{-3i\theta})xyz\\
-(ax+by+cz)yz-(cx+ay+bz)zx
-(bx+cy+az)xy.
\end{aligned}
$$
First of all, we check that all the $2\times 2$ principal minors are nonnegative.
For example, the third  $2\times2$ minor is
$$
M_1:=(cx+ay+bz)(bx+cy+az)-yz,
$$
and we have
$$
M_1\ge \frac1{by+cz}F(0,y,z)
= (ay+bz)(cy+az)-yz
=ac y^2 +(a^2+bc-1)yz+abz^2.
$$
This is nonnegative when $a\ge 1$. If $0\le a\le 1$ then we use the condition (\ref{p2})
to see easily that this quadratic form is nonnegative for each $y,z\ge 0$.
In the same way, we see that all $2\times2$ minors are nonnegative, and $F(x,y,z)\ge 0$ whenever
one of $x,y$ or $z$ is zero.

Now, we show that $F(x,y,z)\ge 0$ on the region  $\{(x,y,z): x,y,z>0\}$.
First, we note that all of $\dfrac{\partial F}{\partial x}$, $\dfrac{\partial F}{\partial y}$
and $\dfrac{\partial F}{\partial z}$ are quadratic
forms associated with the following symmetric matrices;
$$
\left(\begin{matrix}P&R&Q\\R&Q&S\\Q&S&R\end{matrix}\right),\qquad
\left(\begin{matrix}R&Q&S\\Q&P&R\\S&R&Q\end{matrix}\right),\qquad
\left(\begin{matrix}Q&S&R\\S&R&Q\\R&Q&P\end{matrix}\right),\qquad
$$
where
$$
\begin{aligned}
P&=3abc,\\
Q&=a^2c+b^2a+c^2b-c,\\
R&=a^2b+b^2c+c^2a-b,\\
2S&=a^3+b^3+c^3+3abc-3a-(e^{3i\theta}+e^{-3i\theta}).
\end{aligned}
$$
That is, $\dfrac{\partial F}{\partial x}(x,y,z)$ is expressed by
\[
\dfrac{\partial F}{\partial x}(x,y,z)=
\begin{pmatrix} x & y & z \end{pmatrix}
\begin{pmatrix} P&R&Q\\R&Q&S\\Q&S&R \end{pmatrix}
\begin{pmatrix} x\\y\\z\end{pmatrix},
\]
for example. In the case of $a\ge 1$, we have
$$
P+Q+R\ge 3bc+(b^2+c^2)+bc(b+c)\ge 0,
$$
and the equality holds if and only $b=c=0$.
In the case of $0\le a <1$, we have $bc\ge (1-a)^2$ by (\ref{p2}). Hence, it follows that
$$
\begin{aligned}
P+Q+R &=abc+(b+c)(a^2+a(b+c)+bc-1)\\
&\ge abc+(b+c)(a^2+2a(1-a)+(1-a)^2-1)=abc,
\end{aligned}
$$
and the equality holds if and only if $a=0,\,bc=1$.
Consequently, we have $P+Q+R\ge 0$ for all cases.

First, we consider the case of $P+Q+R=0$, from which we have the following two cases:
\begin{itemize}
\item $a=0,\, bc=1$,
\item $a\ge 1,\, b=c=0$.
\end{itemize}
In the first case, we already know that the map is completely copositive.
In the second case, we have $a\ge p_{\theta}$ by (\ref{p1}), and so the map is completely positive.
Therefore, if $P+Q+R=0$ then we see that $\Phi[a,b,c;\theta]$ is positive.

Now, we assume that $P+Q+R>0$. In this case, we have $abc\neq 0$ except for the following two cases:
\begin{itemize}
\item $ab\neq 0$ and $c=0$,
\item $ac\neq 0$ and $b=0$.
\end{itemize}
For the case of $ab\neq 0$ and $c=0$, we have $a\ge 1$ from the condition \eqref{p2}, and so
\begin{equation*}
\begin{aligned}
F(x,y,z)
=&b(a^2-1)(x^2 y+y^2 z+z^2 x)+ab^2(x^2 z+y^2 x+z^2 y)\\
&\phantom{ZZZZZZZZZZZZZ}+(a^3+b^3-3a-(e^{3i\theta}+e^{-3i\theta}))xyz\\
\ge &(3b(a^2-1)+3ab^2+a^3+b^3-3a-(e^{3i\theta}+e^{-3i\theta}))xyz\\
=&((a+b)^3-3(a+b)-(e^{3i\theta}+e^{-3i\theta}))xyz,
\end{aligned}
\end{equation*}
which is nonnegative by the condition \eqref{p1}.
Similarly, one can show that $F(x,y,z)$ is nonnegative for the case of $ac\neq 0$ and $b=0$.

Now, we consider the case of $abc\neq 0$.
In this case, the coefficients of $x^3,y^3$ and $z^3$ in the polynomial $F$ are positive, and so
there exists a sufficiently large cube
$R=\{(x,y,z): 0\le x,y,z\le M\}$
so that  $F(x,y,z)\ge 0$ outside of $R$. Furthermore, we already know that $F(x,y,z)\ge 0$ if $xyz=0$.
Therefore, it suffices to show that the local minimums of $F$ in the region  $\{(x,y,z): x,y,z>0\}$
are nonnegative.

If $(x,y,z)$ is a nontrivial common solution of $\dfrac{\partial F}{\partial x}$,
$\dfrac{\partial F}{\partial y}$ and $\dfrac{\partial F}{\partial z}$ then
it is also a nontrivial solution of homogeneous quadratic equation given by
\begin{equation}\label{quadratic_form}
\left(\begin{matrix}P+Q+R&Q+R+S&Q+R+S\\Q+R+S&P+Q+R&Q+R+S\\Q+R+S&Q+R+S&P+Q+R\end{matrix}\right).
\end{equation}
This means that a nontrivial common solution $(x,y,z)$ of $\dfrac{\partial F}{\partial x}$,
$\dfrac{\partial F}{\partial y}$ and $\dfrac{\partial F}{\partial z}$ satisfies
\begin{equation}\label{expansion}
\begin{aligned}
&\begin{pmatrix} x & y & z\end{pmatrix}
\begin{pmatrix}P+Q+R&Q+R+S&Q+R+S\\Q+R+S&P+Q+R&Q+R+S\\Q+R+S&Q+R+S&P+Q+R\end{pmatrix}
\begin{pmatrix} x\\y\\z \end{pmatrix}\\
=& (P+Q+R)(x^2+y^2+z^2)+2(Q+R+S)(xy+yz+zx)=0.
\end{aligned}
\end{equation}
If $P=S$ then common solutions are on the plane $x+y+z=0$, and so
there is no nonzero common solution in the region $\{(x,y,z):x,y,z>0\}$. Therefore, we may assume $P\neq S$. In this case, the $2\times 2$ minors is not zero, and so the rank of the matrix~\eqref{quadratic_form} is more than $2$.
We also note that the determinant of this matrix~\eqref{quadratic_form} is
$$
(P-S)^2(3Q+3R+P+2S)=(P-S)^2[(a+b+c)^3-3(a+b+c)-(e^{3i\theta}+e^{-3i\theta})]\ge 0
$$
by the condition (\ref{p1}). If $P> S$ then we see that the above matrix is positive semi-definite, and so
it must be singular with rank two. Therefore, a common solution must belong to the $1$-dimensional kernel space of the matrix~\eqref{quadratic_form}. Consequently, all common solutions are of the form $(x,x,x)$. We consider the case $P<S$. In this case, we see that common solution satisfies
$$
\begin{aligned}
& (P+Q+R)(x^2+y^2+z^2)+2(Q+R+S)(xy+yz+zx)\\
=&(P+Q+R)(x+y+z)^2+2(S-P)(xy+yz+zx)=0
\end{aligned}
$$
which is impossible in the region $\{(x,y,z):x,y,z>0\}$.
Summing up, we see that if $F$ takes a local minimum at $(x,y,z)$ with $x,y,z>0$ then $x=y=z$.
We note that
$$
\begin{aligned}
\frac 1{x^3}F(x,x,x)
&=
a^3 + b^3 + c^3 + 3(a^2b+a^2c+b^2c+b^2a+c^2a+c^2b+2abc)\\
&\phantom{ZZZZZZZZ}-(e^{3i\theta}+e^{-3i\theta})-3(a+b+c)\\
&=(a+b+c)^3-3(a+b+c)-(e^{3i\theta}+e^{-3i\theta}),
\end{aligned}
$$
which is nonnegative by (\ref{p1}) when $x\neq 0$.
 This completes the proof for the following:
\begin{theorem}
The map $\Phi[a,b,c;\theta]$ is positive if and only if both conditions {\rm (\ref{p1})} and {\rm (\ref{p2})} hold.
\end{theorem}

\section{Facial structures}

For each $\theta$, we denote by $\Gamma^\theta$ the $3$-dimensional convex body determined by (\ref{p1}) and (\ref{p2}).
The facial structures of the convex body $\Gamma^0$ has been analyzed in \cite{ha_kye_opt_ind} for the case of $\theta=0$.
Facial structures of $\Gamma^\theta$ is similar as those of $\Gamma^0$ except several differences.
We first consider the case $1<p_\theta\le 2$. In this case, the convex body $\Gamma^\theta$ has the following four $2$-dimensional faces:
\begin{itemize}
\item
$f^\theta_{\rm ab}=\{(a,b,c): c=0,\ a+b\ge p_\theta,\ a\ge 1\}$,
\item
$f^\theta_{\rm ac}=\{(a,b,c): b=0,\ a+c\ge p_\theta,\ a\ge 1\}$,
\item
$f^\theta_{\rm bc}=\{(a,b,c): a=0,\ bc\ge 1\}$,
\item
$f^\theta_{\rm abc}=\{(a,b,c): a+b+c=p_\theta,\ 0\le a\le 1\Longrightarrow bc\ge (1-a)^2\}$.
\end{itemize}
\begin{figure}[h!]
\begin{center}
\includegraphics[scale=0.8]{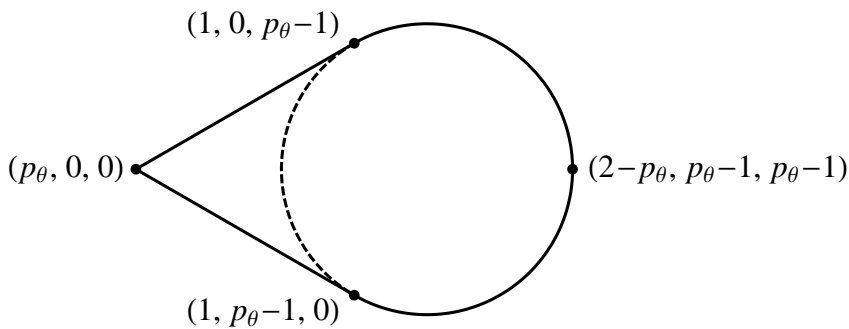}
\end{center}
\caption{Picture of the face $f^\theta_{\rm abc}$ in the plane $a+b+c=p_{\theta}$.}
\end{figure}
In the case of $p_\theta=1$, the face $f^\theta_{\rm abc}$ shrinks to a single point $(1,0,0)$.
In order to figure out the shape of the face $f^\theta_{\rm abc}$, we modify the parametrization in \cite{ha+kye_indec-witness} to put
\begin{equation}\label{para}
\begin{aligned}
a_\theta(t)&=(p_\theta-1)\cdot\dfrac{(1-t)^2}{1-t+t^2}+(2-p_\theta)=1-\dfrac {(p_\theta-1)t}{1-t+t^2},\\
b_\theta(t)&=(p_\theta-1)\cdot\dfrac{t^2}{1-t+t^2},\\
c_\theta(t)&=(p_\theta-1)\cdot\dfrac{1}{1-t+t^2},
\end{aligned}
\end{equation}
for $0< t<\infty$. 
Then we have
$$
a_\theta(t)+b_\theta(t)+c_\theta(t)=p_\theta,\qquad 0\le a_\theta(t)\le 1,\qquad b_\theta(t)c_\theta(t)=(1-a_\theta(t))^2.
$$
If $p_\theta=2$ then this face touches the face $f_{\rm bc}$ at the point $(0,1,1)$ which gives rise to a completely copositive map.
On the other hand, if $1<p_\theta<2$ then this face does not touch the face $f_{\rm bc}$.

The convex body $\Gamma^\theta$ also has the following $1$-dimensional faces:
\begin{itemize}
\item
$e^\theta_{\rm a}=\{(a,0,0): a\ge p_\theta\}$,
\item
$e^\theta_{\rm b}=\{(1,b,0): b\ge p_{\theta}-1\}$,
\item
$e^\theta_{\rm c}=\{(1,0,c): c\ge p_{\theta}-1\}$,
\item
$e^\theta_{\rm ab}=\{(a,b,0): a+b=p_\theta,\, 1\le a \le p_\theta\}$,
\item
$e^\theta_{\rm ac}=\{(a,0,c): a+c=p_\theta,\, 1\le a \le p_\theta\}$,
\item
$\displaystyle{   e^\theta_t=\left\{\left(1-s,st,\frac st\right): \dfrac{(p_\theta-1)t}{1-t+t^2}\le s\le 1\right\}   }$ for $t>0$.
\end{itemize}
We note that $e^\theta_t$ is the line segment from the point $(a_\theta(t),b_\theta(t),c_\theta(t))$ to the point $(0,t,\frac 1t)$,
and lies on the surface $bc= (1-a)^2$ for $0\le a < 1$.
We also note that $e^\theta_1$ shrink to a single point $(0,1,1)$ if $p_\theta=2$.
It remains to list up $0$-dimensional faces as follows:
\begin{itemize}
\item
$v_{(p_\theta,0,0)}$,
\item
$v_{(1,0,p_\theta-1)},\ v_{(1,p_\theta-1,0)}$,
\item
$v_{(a_\theta(t),b_\theta(t),c_\theta(t))}$ for $t>0$,
\item
$v_{(0,t,1/t)}$ for $t>0$.
\end{itemize}

It $p_\theta=1$ then all of the following faces
$$
f^\theta_{\rm abc},\quad e^\theta_{\rm ab},\quad e^\theta_{\rm ac}, \quad
v_{(p_\theta,0,0)},\quad v_{(1,0,p_\theta-1)},\quad v_{(1,p_\theta-1,0)},\quad v_{(a_\theta(t),b_\theta(t),c_\theta(t))}
$$
shrink to the single point $(1,0,0)$. Furthermore, the face $e^\theta_t$ connects the two points $(1,0,0)$ and $(0,t,\frac 1t)$
which represent completely positive and completely copositive maps, respectively. Therefore, every positive linear map
$\Phi[a,b,c;\theta]$ is decomposable. Since we are interested in indecomposable cases, we assume $p_\theta>1$
throughout this note.

By the exactly same argument as in \cite{ha_kye_opt_ind}, we have the following:
\begin{itemize}
\item
Interior points of $f^\theta_{\rm ab}$, $f^\theta_{\rm ac}$, $f^\theta_{\rm bc}$,
$e^\theta_{\rm a}$, $e^\theta_{\rm b}$ and $e^\theta_{\rm c}$ are neither
optimal nor co-optimal.
\item
Interior points of $f^\theta_{\rm abc}$, $e^\theta_{\rm ab}$, $e^\theta_{\rm ac}$, $v_{(p_\theta,0,0)}$ are not optimal.
\item
Interior points of $e^\theta_t$ and $v_{(0,t,1/t)}$ are not co-optimal.
\item
$v_{(1,0,p_\theta-1)}$ and $v_{(1,p_\theta-1,0)}$ are not spanning.
\end{itemize}
We recall that if two positive map $\phi_1$ and $\phi_2$ determine a common smallest face containing them,
then they are interior points of the common face, and share the above properties related with the optimality.

\begin{figure}[h!]
\includegraphics[scale=0.43]{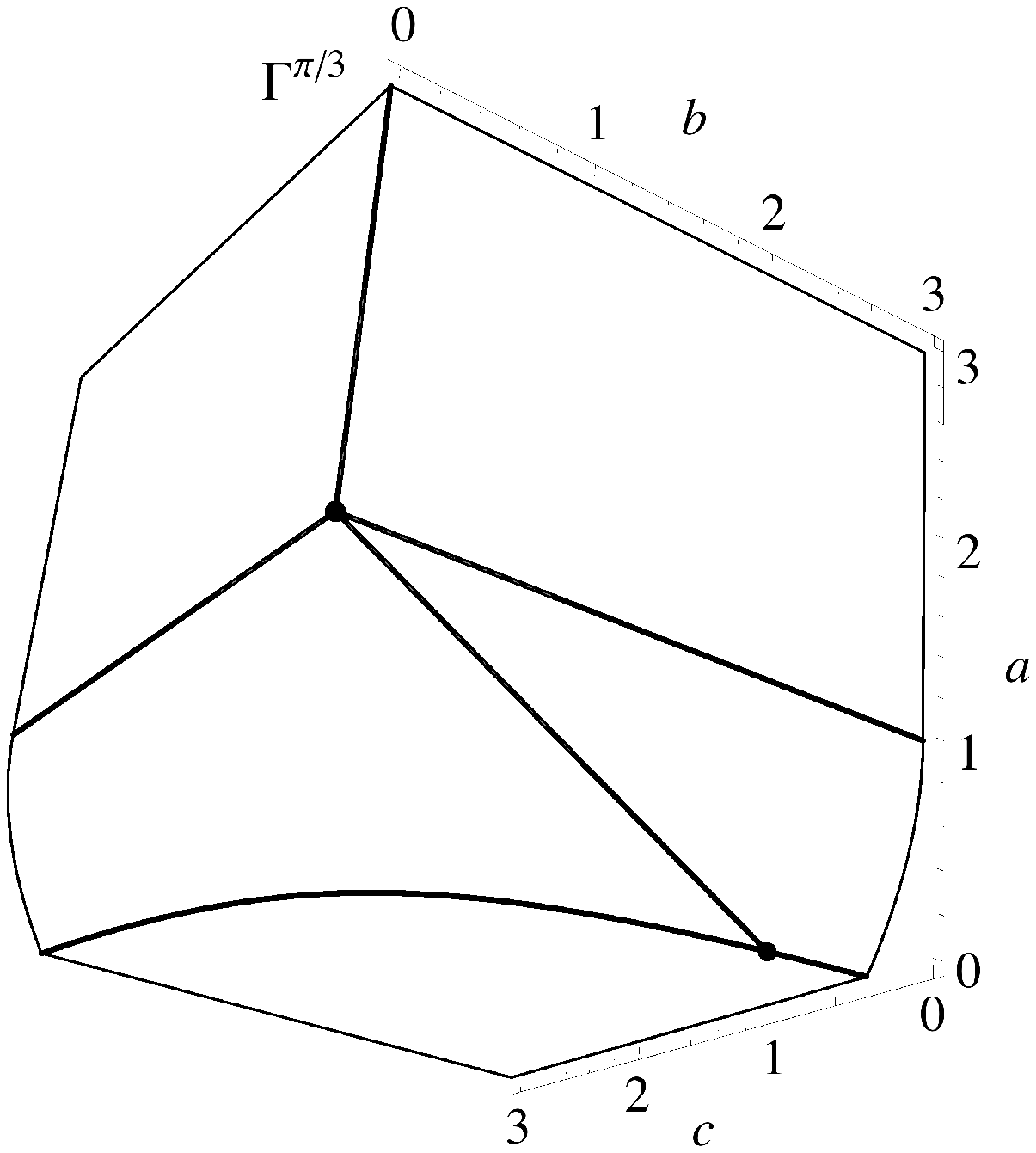}\,
\includegraphics[scale=0.43]{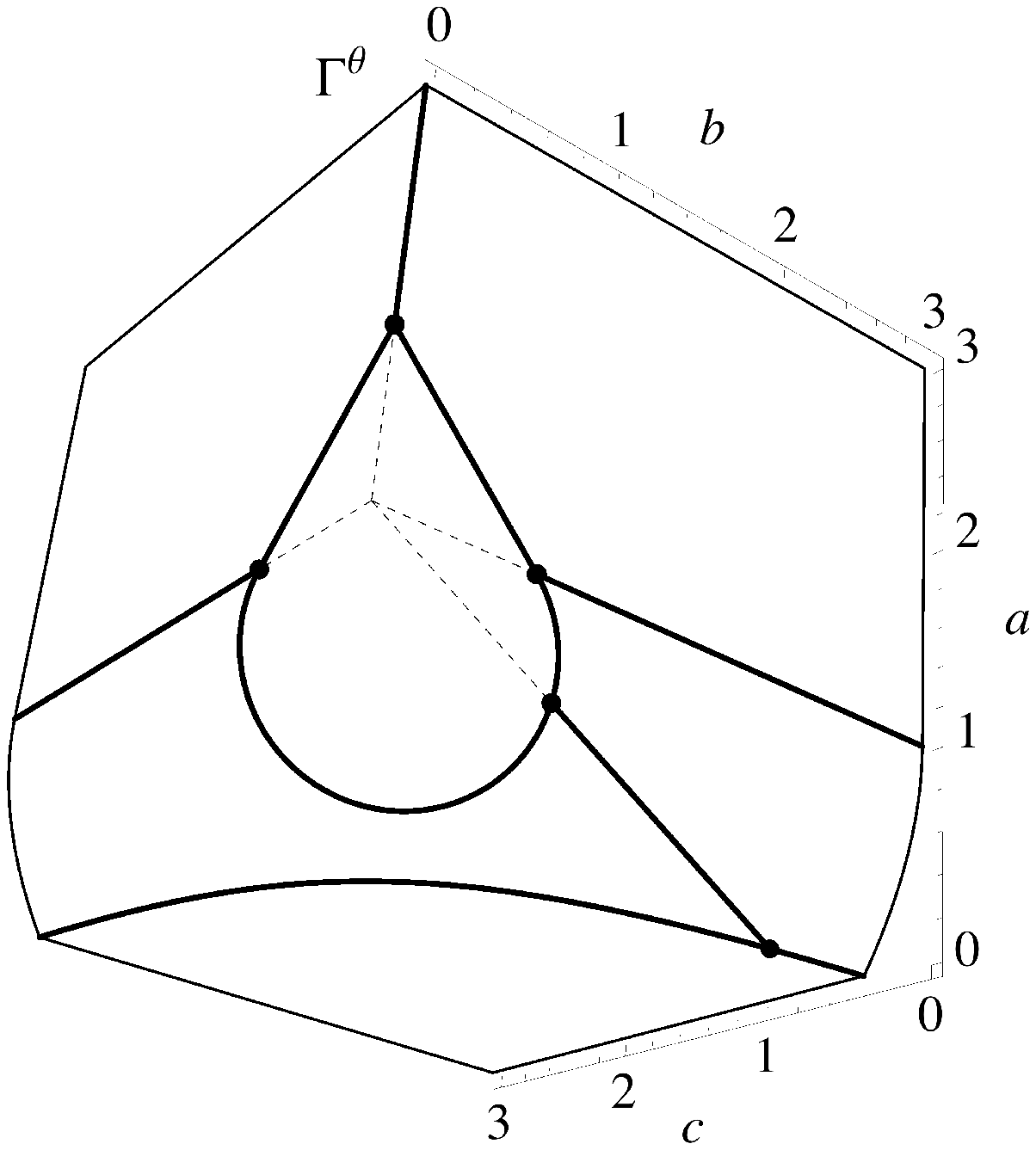}\,
\includegraphics[scale=0.43]{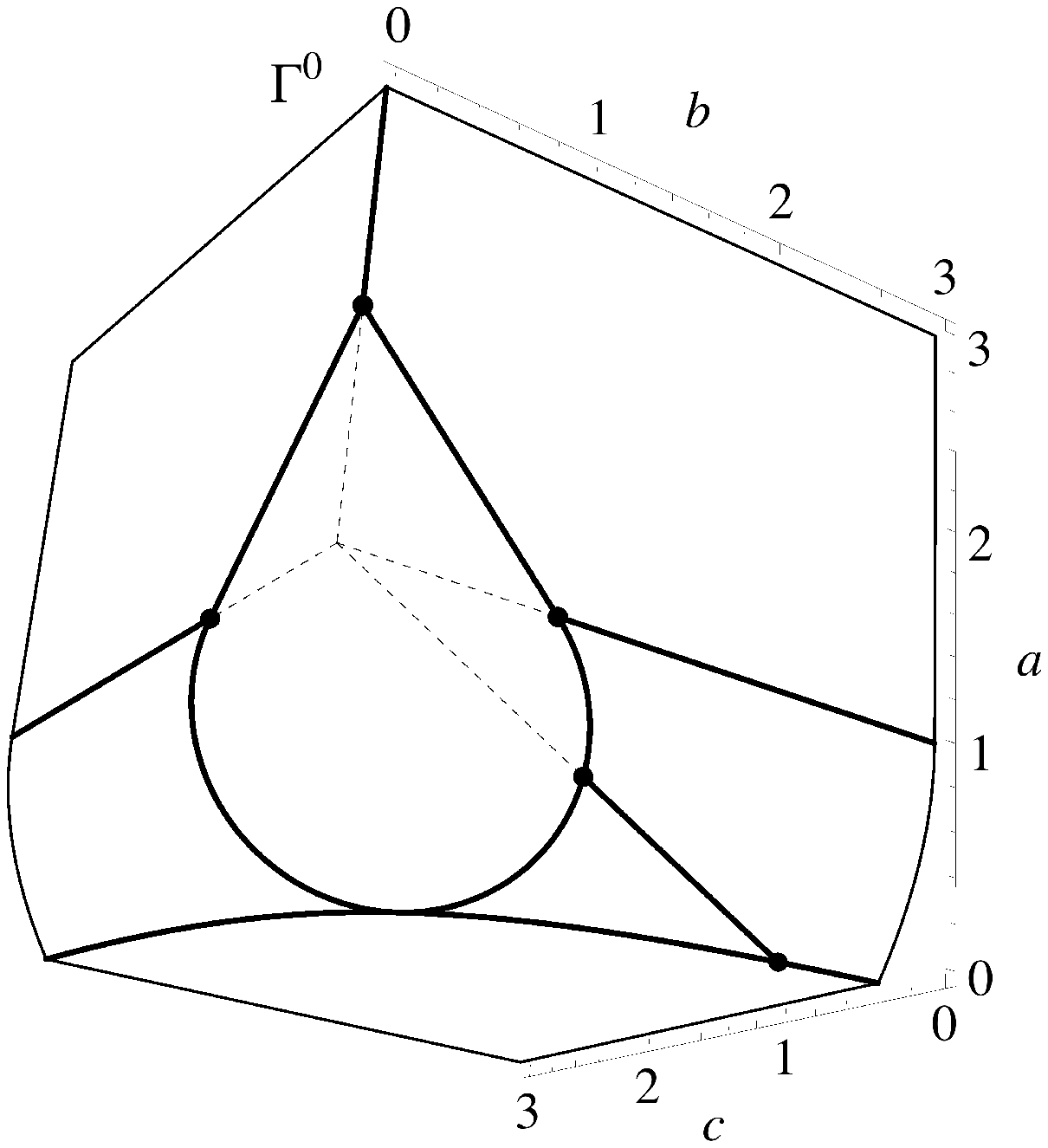}\,
\caption{Figures of the convex bodies for $p_{\theta}=1,\,1<p_{\theta}<2$ and $p_{\theta}=2$.}
\end{figure}

\section{Spanning Properties}

In this section, we determine which positive linear maps have the spanning property and/or the co-spanning property.
We remind the readers that we are assuming that $p_\theta>1$. We also assume that $p_\theta<2$,
since the case of $\theta=0$ is already considered in \cite{ha_kye_opt_ind}.
We note that the spanning property (respective co-spanning property) implies the optimality (respectively co-optimality).
By the discussion of the previous section, it remains to consider the following cases:
\begin{enumerate}
\item[(i)]
$0< a\le 1,\ bc=(1-a)^2,\ a+b+c> p_\theta$,
\item[(ii)]
$2-p_\theta\le a\le 1,\ bc=(1-a)^2,\ a+b+c= p_\theta$,
\item[(iii)]
$a=0, \ bc= 1$,
\item[(iv)]
$1<a\le p_\theta,\ a+b+c= p_\theta$.
\end{enumerate}

We recall \cite{kye_ritsu} that $\phi\in\mathbb P_1$ has the spanning property if and only if the set
$$
P[\phi]:=\{z=\xi\ot\eta:\lan zz^*,\phi\ran=0\}
$$
spans the whole space $\mathbb C^m\ot\mathbb C^n$, and
$$
\lan zz^*,\phi\ran
=\lan \xi\xi^*\ot \eta\eta^*,\phi\ran
=\tr(\phi(\xi\xi^*)\bar \eta\bar \eta^*)
=(\phi(\xi\xi^*)\bar \eta|\bar \eta).
$$
Therefore, we see that $\xi\ot\eta\in P[\phi]$ if and only if $\phi(\xi\xi^*)\bar\eta=0$.
In order to determine the set $P[\Phi[a,b,c;\theta]]$, we first
find vectors $(x,y,z)\in\mathbb C^3$ such that the matrix (\ref{mat}) is singular. In other words, we look for $(x,y,z)$
for which $F(x,y,z)=0$. The only possibility of $F(x,y,z)=0$ with nonzero $x,y,z$ is $F(x,x,x)=0$, and this happens
only when the equality holds in (\ref{p1}).

Now, we consider the case (i). In this case, we see that $F(x,y,z)=0$
holds only if $xyz=0$. We first consider the case $z=0$, for which we have
$$
F(x,y,z)
=(acx^2+(a^2+bc-1)xy+aby^2)(bx+cy)
=a(\sqrt cx-\sqrt by)^2(bx+cy).
$$
Therefore, the matrix (\ref{mat}) is singular with $z=0$ if and only if $(x,y,0)=(b^{1/4}\alpha,c^{1/4}\beta,0)$ with
complex numbers $\alpha,\beta$ with modulus one. In this case, (\ref{mat}) is given by
$$
\left(\begin{matrix}
\sqrt b &-e^{i\theta}\alpha\bar\beta\sqrt{1-a} &0\\
-e^{-i\theta}\bar\alpha\beta\sqrt{1-a}&\sqrt c&0\\
0&0&b\sqrt b+c\sqrt c
\end{matrix}\right),
$$
and the kernel is $(\alpha e^{i\theta}\sqrt{1-a},\beta\sqrt b,0)$. In the same way, we see that $z$
belongs to $P[\Phi[a,b,c;\theta]]$  if and only if $z$ is one of the following:
\begin{equation}\label{prod}
\begin{aligned}
z_1[\alpha,\beta]=&(b^{1/4}\alpha,c^{1/4}\beta,0)^{\rm t}\otimes (\bar{\alpha} e^{-i\theta}\sqrt{1-a},\bar{\beta}\sqrt b,0)^{\rm t},\\
z_2[\alpha,\beta]=&(c^{1/4}\beta,0,b^{1/4}\alpha)^{\rm t}\otimes (\bar{\beta}\sqrt b,0,\bar{\alpha} e^{-i\theta}\sqrt{1-a})^{\rm t},\\
z_3[\alpha,\beta]=&(0,b^{1/4} \alpha,c^{1/4}\beta)^{\rm t}\otimes (0,\bar{\alpha} e^{-i\theta}\sqrt{1-a},\bar{\beta}\sqrt b)^{\rm t},
\end{aligned}
\end{equation}
with complex numbers $\alpha,\,\beta$ with modulus one.
It is clear that these vectors do not span the whole space if $a=1$ which implies $bc=0$ in this case.

We consider the case $0<a<1$.
We take $\beta_1=1,\, \beta_2=-1,$ and
$\beta_3=i$, and consider the $9\times 9$ matrix whose columns are
nine vectors $z_k[1,\beta_\ell]$ for $k,\,\ell=1,2,3$. Then the
determinant of $M$ is given by
$$
|\det M|=|64b^{\frac 92}c^{\frac 94}e^{-3i \theta} i (1+e^{-3i\theta})|
$$
which is nonzero, since $\theta\neq\pm\frac\pi 3,\pm\pi$, and $a<1$
implies that $bc\neq 0$. Therefore, we conclude that
$\Phi[a,b,c;\theta]$ has the spanning property if and only if $a<1$
for the case (i). It is clear that it has not the co-spanning
property from the facial structures in the previous section.

Now, we consider the case (ii). First of all, we note that product
vectors in (\ref{prod}) already belong to $P[\Phi[a,b,c;\theta]]$,
and so we see that $\Phi[a,b,c;\theta]$  has the spanning property
if $a<1$. We will see in the next section that
$\Phi[1,0,p_\theta-1;\theta]$ and $\Phi[1,p_\theta-1,0;\theta]$ are
optimal, but do not have the spanning property. We look for another
product vectors in $P[\Phi[a,b,c;\theta]]$ to determine if they have
the co-spanning property. If $x=(x_1,x_2,x_3)^{\rm t}$ with
$|x_1|=|x_2|=|x_3|$, then $\Phi[a,b,c;\theta](xx^*)$ is given by
\[
\begin{pmatrix} |x_1|^2 p_{\theta} & -e^{i\theta} x_1\bar{x}_2 & -e^{-i\theta}x_1\bar{x}_3\\
-e^{-i\theta}x_2\bar{x}_1 & |x_1|^2 p_{\theta} & -e^{i\theta} x_2\bar{x}_3\\
-e^{i\theta}x_3\bar{x}_1 & -e^{-i\theta}x_3 \bar{x}_2& |x_1|^2 p_{\theta}
\end{pmatrix},
\]
for which
\begin{itemize}
\item
$(x_1,x_2e^{i\frac 23\pi },x_3e^{-i\frac 23\pi })^{\rm t}$ is a kernel vector if $-\pi \le \theta \le -\frac \pi 3$.
\item
$(x_1,x_2,x_3)^{\rm t}$ is a kernel vector if $-\frac \pi 3\le\theta\le\frac \pi 3$.
\item
$(x_1,x_2e^{-i\frac 23\pi },x_3e^{i\frac 23\pi })^{\rm t}$ is a kernel vector if $\frac \pi 3\le\theta\le\pi$.
\end{itemize}
Therefore, we see that $z\in P[\Phi[a,b,c;\theta]]$  if and only if
$z$ is either one of the vectors in \eqref{prod} or one of the following form:
\begin{equation}\label{prod2}
\begin{aligned}
w[\alpha,\beta,\gamma]=&
   (\alpha,\beta,\gamma)^{\rm t}\otimes (\bar{\alpha} ,\bar{\beta}e^{-i \frac{2}3\pi},\bar{\gamma}e^{i \frac{2}3\pi})^{\rm t}
   \quad \text{ if }{\textstyle -\pi < \theta < -\frac \pi 3}, \\
w[\alpha,\beta,\gamma]=&
   (\alpha,\beta,\gamma)^{\rm t}\otimes (\bar{\alpha},\bar{\beta},\bar{\gamma})^{\rm t}
   \quad \text{ if }{\textstyle -\frac \pi 3<\theta<\frac \pi 3},\\
w[\alpha,\beta,\gamma]=&
   (\alpha,\beta,\gamma)^{\rm t}\otimes (\bar{\alpha} ,\bar{\beta}e^{i \frac{2}3\pi},\bar{\gamma}e^{-i \frac{2}3\pi})^{\rm t}
   \quad \text{ if }{\textstyle \frac \pi 3<\theta<\pi}, \\
\end{aligned}
\end{equation}
with $|\alpha|=|\beta|=|\gamma|$.
Now, we take product vectors in $P[\Phi[a,b,c;\theta]]$ as follows:
\begin{equation}\label{cond2_prod1}
z_1[1,1],\,z_1[1,-1],\,z_2[1,1],\,z_2[1,-1],\,z_3[1,1],\,z_3[1,-1],\,w[1,1,1],\,w[1,-1,1],\,w[1,i,-i].
\end{equation}
We consider the $9\times 9$ matrix whose columns the partial
conjugates of the above nine vectors, then the determinant is given
as follows:
\begin{equation*}
|\det M|=\begin{cases}
16\sqrt{2}b^{\frac 94}c^{\frac 34}|(\sqrt{b}-\sqrt{c}e^{i(\theta+\frac 23 \pi)})^3(1+e^{3i\theta})|
  & \text{ if\ \ } {\textstyle -\pi < \theta < -\frac \pi 3},\\
16\sqrt{2}b^{\frac 94}c^{\frac 34}|(\sqrt{b}-\sqrt{c}e^{i\theta})^3(1+e^{3i\theta})|
  & \text{ if \ \ } {\textstyle -\frac \pi 3<\theta<\frac \pi 3},\\
16\sqrt{2}b^{\frac 94}c^{\frac 34}|(\sqrt{b}-\sqrt{c}e^{i(\theta-\frac 23 \pi)})^3(1+e^{3i\theta})|
  & \text{ if \ \ } {\textstyle \frac \pi 3<\theta<\pi}.\\
\end{cases}
\end{equation*}
We note that
$\det M=0$ implies $b=c$ and $\theta=0,\pm\frac 23\pi$,
which is not possible by the assumption $p_{\theta}<2$.
Therefore, partial conjugates of
the product vectors in \eqref{cond2_prod1} 
span the whole space
$\mathbb C^3\otimes \mathbb C^3$, and $W[a,b,c;\theta]$ has the co-spanning property for the case (ii).

Now, we consider the case (iii). In this case, we note that
$\Phi[a,b,c,\theta]$ is completely copositive, and so they never
satisfies the co-spanning property. Note that the matrix \eqref{mat}
is given by
\[
\begin{pmatrix}
b|y|^2 & -e^{i\theta} x \bar y & 0\\
-e^{-i\theta} y\bar x & c|x|^2 & 0 \\
0 & 0 & b|x|^2+c|y|^2
\end{pmatrix},
\]
and the kernel is $(x,e^{-i\theta} yb,0)^{\rm t}$. Therefore, the following vectors
$$
\begin{aligned}
w_1[\alpha,\beta]=&(\alpha,\beta,0)^{\rm t}\otimes (\bar{\alpha},e^{i\theta}\bar{\beta} b,0)^{\rm t},\\
w_2[\alpha,\beta]=&(\beta,0,\alpha)^{\rm t}\otimes (e^{i\theta}\bar{\beta}b,0,\bar{\alpha})^{\rm t},\\
w_3[\alpha,\beta]=&(0,\alpha,\beta)^{\rm t}\otimes (0,\bar{\alpha},e^{i\theta}\bar{\beta}b)^{\rm t}
\end{aligned}
$$
belong to $P[\Phi[a,b,c;\theta]$. We see that the set of the
following vectors
\[
w_1[1,1],\,w_1[1,-1],\,w_1[1,i],\,w_2[1,1],\,w_2[1,-1],\,w_2[1,i],\,w_3[1,1],\,w_3[1,-1],\,w_3[1,i]
\]
span the whole space $\mathbb C^3\otimes \mathbb C^3$,
because the determinant of $9\times 9$ matrix $M$ whose columns are the above vectors is given by
\[
|\det(M)|=64b^3 |1+b^3 e^{3i\theta}|,
\]
which is nonzero by the assumption $p_\theta>1$.
Therefore, we see that the map $\Phi[a,b,c;\theta]$ has the spanning property
for the case (iii).

It remains to consider the case (iv). In this case, they never have the spanning property, since $\Phi[p_\theta,0,0;\theta]$
is completely positive. We consider interior points of the $2$-dimensional face $f^\theta_{\rm abc}$
on the plane $a+b+c=p_\theta$. In this case,
the only possible product vectors in $P[\Phi[a,b,c;\theta]]$ are of the form (\ref{prod2}). It is clear that the partial conjugate of them
do not span the whole space, and so they do not have the co-spanning property. In the next section, we will see that
they are not co-optimal. Finally, we consider the
line segment $e^\theta_{\rm ac}$ between two points $(p_\theta,0,0)$ and $(1,0,p_\theta-1)$. We note that the smallest exposed face $F$
containing $\Phi[1,0,p_\theta-1;\theta]$ is bigger than $e^\theta_{\rm ac}$. We already shown that $\Phi[1,0,p_\theta-1;\theta]$ has the co-spanning property,
and so $F$ has no completely copositive map. See \cite{choi_kye} for the case of $\theta=0$.
This shows that the line segment $e^\theta_{\rm ac}$ has the co-spanning property.

\begin{theorem}\label{thm:spanning}
Suppose that the map $\Phi[a,b,c;\theta]$ is positive, and $1<p_\theta<2$. Then we have the following:
\begin{enumerate}
\item[(i)]
$\Phi[a,b,c;\theta]$ has the spanning property if and only if
$$
0\le a<1,\quad bc=(1-a)^2.
$$
\item[(ii)]
$\Phi[a,b,c;\theta]$ has the co-spanning property if and only if
$$
2-p_\theta\le a\le 1,\quad bc=(1-a)^2,\quad a+b+c= p_\theta
$$
holds or
$$
1\le a\le p_\theta,\quad  bc=0,\quad a+b+c= p_\theta.
$$
\end{enumerate}
\end{theorem}

We summarize the results in terms of faces for $1<p_\theta<2$ as follows:
\begin{itemize}
\item
$e^\theta_t$, $v_{(a_\theta(t),b_\theta(t),c_\theta(t))}$ and $v_{(0,t,1/t)}$ have the spanning property.
\item
$v_{(1,0,p_\theta-1)}$ and $v_{(1,p_\theta-1,0)}$ have not the spanning property. It should be checked if they are optimal or not.
\item
$e^\theta_{\rm ab}$, $e^\theta_{\rm ac}$, $v_{(p_\theta,0,0)}$,
$v_{(1,0,p_\theta-1)}$, $v_{(1,p_\theta-1,0)}$ and $v_{(a_\theta(t),b_\theta(t),c_\theta(t))}$ have the co-spanning property.
\item
$f^\theta_{\rm abc}$ have not the co-spanning property. It should be checked if it is co-optimal or not.
\end{itemize}

\section{Optimality without the spanning property}

It remains to check the optimality of $\Phi[1,0,p_\theta-1;\theta]$ and $\Phi[1,p_\theta-1,0;\theta]$, and co-optimality
of interior points of the face $f^\theta_{\rm abc}$.
We recall that $\Phi[1,0,1;0]$ and $\Phi[1,1,0;0]$, which are usually called the Choi map, are extremal when $\theta=0$ by \cite{choi-lam},
and so they are optimal.
In order to check the optimality of a positive map $\phi$,
we first find all extremal completely positive maps $\phi_V$ in the smallest
exposed face determined by $\phi$, and check if they belong to the smallest face determined by $\phi$.
Recall that $\phi_V$ is the completely positive map given by
$$
\phi_V(X)=V^*XV,\qquad X\in M_m,
$$
where $V$ is an $m\times n$ matrix.
We also recall \cite{kye_ritsu} that $\phi_V$ belongs to the smallest exposed face determined by $\phi$ if and only if
$V$ is orthogonal to $P[\phi]$ when we identify an $m\times n$ matrix and a vector
in $\mathbb C^m\otimes \mathbb C^n$.

We proceed to show that  $\Phi[1,p_\theta-1,0;\theta]$ is optimal. We first consider the case $-\frac\pi 3<\theta<\frac\pi 3$.
In this case, $P[\Phi[1,p_\theta-1,0;\theta]]$ consists of
$$
e_1\ot e_2,\quad
e_2\ot e_3,\quad
e_3\ot e_1,\quad
(\alpha,\beta,\gamma)^{\rm t}\otimes (\bar{\alpha},\bar{\beta},\bar{\gamma})^{\rm t},
$$
with complex numbers $\alpha,\beta$ and $\gamma$ with modulus one, from
(\ref{prod}) and (\ref{prod2}). Every vector orthogonal to all of these vectors is of the form
$$
(\xi,0,\,0;\, 0,\eta,0,\,;\, 0,0,\zeta)^\ttt, \qquad \xi+\eta+\zeta=0,
$$
and the Choi matrix of the completely positive map associated with this vector is given by
$$
V[\xi,\eta,\zeta]
=\left(
\begin{array}{ccccccccccc}
|\xi|^2     &\cdot   &\cdot  &\cdot  &\xi\bar\eta     &\cdot   &\cdot   &\cdot  &\xi\bar\zeta     \\
\cdot   &\cdot &\cdot    &\cdot    &\cdot   &\cdot &\cdot &\cdot     &\cdot   \\
\cdot  &\cdot    &\cdot &\cdot &\cdot  &\cdot    &\cdot    &\cdot &\cdot  \\
\cdot  &\cdot    &\cdot &\cdot &\cdot  &\cdot    &\cdot    &\cdot &\cdot  \\
\eta\bar\xi     &\cdot   &\cdot  &\cdot  &|\eta|^2     &\cdot   &\cdot   &\cdot  &\eta\bar\zeta    \\
\cdot   &\cdot &\cdot    &\cdot    &\cdot   &\cdot &\cdot &\cdot    &\cdot   \\
\cdot   &\cdot &\cdot    &\cdot    &\cdot   &\cdot &\cdot &\cdot    &\cdot   \\
\cdot  &\cdot    &\cdot &\cdot &\cdot  &\cdot    &\cdot    &\cdot &\cdot  \\
\zeta\bar\xi     &\cdot   &\cdot  &\cdot  &\zeta\bar\eta     &\cdot   &\cdot   &\cdot  &|\zeta|^2
\end{array}
\right).
$$
In order to show the optimality of $\Phi[1,p_\theta-1,0;\theta]$, we show that if
$$
W[1,p_\theta-1,0;\theta]-pV[\xi,\eta,\zeta]
$$
is block-positive for a nonnegative $p>0$ with $\xi+\eta+\zeta=0$ then $\xi=\eta=\zeta=0$. First, we take product vectors
$z_t=(\sqrt t e^{-i\theta},t,0)^{\rm t}\otimes (\sqrt t,1,0)$ for $t>0$, then we have
\[
\lan z_tz_t^*,W[1,p_\theta-1,0;\theta]- pV[\xi,\eta,\zeta]\ran
=t^2\left( t(p_{\theta}-1)-p|\xi e^{-i\theta}+\eta|^2\right)\ge 0
\]
for all $t>0$ if and only if $\eta =-e^{-i\theta} \xi$. This implies
$\zeta=(e^{-i\theta}-1)\xi$ from the relation $\xi+\eta+\zeta=0$.
Now, we take product vectors $w_t=(0,\sqrt t e^{-i\theta},t)^{\rm
t}\otimes (0,\sqrt t,1)$ for $t>0$. Then we have
\[
\lan w_tw_t^*,W[1,p_\theta-1,0;\theta]- pV[\xi,-e^{-i\theta}\xi,(e^{-i\theta}-1)\xi]\ran
=t^2\left( t(p_{\theta}-1)-p|\xi|^2(1-2\cos \theta)^2\right)\ge 0
\]
for all $t>0$ if and only if $|\xi|^2(1-2\cos\theta)^2=0$. Since
$(1-2\cos \theta)\neq 0$ for $|\theta|<\pi/3$,
we conclude that $\xi=0$. 
Consequently, we have $\xi=\eta=\zeta=0$, and this completes the proof of the optimality of $\Phi[1,p_\theta-1,0;\theta]$.
For another ranges of $\theta$, we have the similar argument.

Now, we show that an interior point $\Phi[1,b,c;\theta]$, with $b+c=p_\theta-1$, of the face $f^\theta_{\rm abc}$ is not co-optimal.
In the case of $\theta=0$, it is clear.
We first consider the case $-\frac\pi 3<\theta<\frac\pi 3\, (\theta\neq 0)$.
To see this, we consider the completely copositive linear map whose
Choi matrix is given by
$$
W[0,1,1;0]
=\left(
\begin{array}{ccccccccccc}
\cdot     &\cdot   &\cdot  &\cdot  &-1     &\cdot   &\cdot   &\cdot  &-1     \\
\cdot   &1 &\cdot    &\cdot    &\cdot   &\cdot &\cdot &\cdot     &\cdot   \\
\cdot  &\cdot    &1 &\cdot &\cdot  &\cdot    &\cdot    &\cdot &\cdot  \\
\cdot  &\cdot    &\cdot &1 &\cdot  &\cdot    &\cdot    &\cdot &\cdot  \\
-1     &\cdot   &\cdot  &\cdot  &\cdot     &\cdot   &\cdot   &\cdot  &-1    \\
\cdot   &\cdot &\cdot    &\cdot    &\cdot   &1 &\cdot &\cdot    &\cdot   \\
\cdot   &\cdot &\cdot    &\cdot    &\cdot   &\cdot &1 &\cdot    &\cdot   \\
\cdot  &\cdot    &\cdot &\cdot &\cdot  &\cdot    &\cdot    &1 &\cdot  \\
-1     &\cdot   &\cdot  &\cdot  &-1     &\cdot   &\cdot   &\cdot  &\cdot
\end{array}
\right),
$$
and look for a small positive number $p>0$ so that $W[1,b,c;\theta]-pW[0,1,1;0]$ is block-positive. We note that
$$
W[1,b,c;\theta]-pW[0,1,1;0]
=|e^{i\theta}-p|W\left[\frac 1{|e^{i\theta}-p|},\frac{b-p}{|e^{i\theta}-p|},\frac{c-p}{|e^{i\theta}-p|};\theta^\prime\right],
$$
where $\theta^\prime$ is the argument of $e^{i\theta}-p$.
 First, we take the positive  number $t_0$ so that $e^{i\theta}-t_0=|e^{i\theta}-t_0|e^{\pm i \frac{\pi}3}$, and then pick a positive number $p<t_0$.
\begin{figure}[h!]\label{fig:find_p}
\includegraphics[scale=0.6]{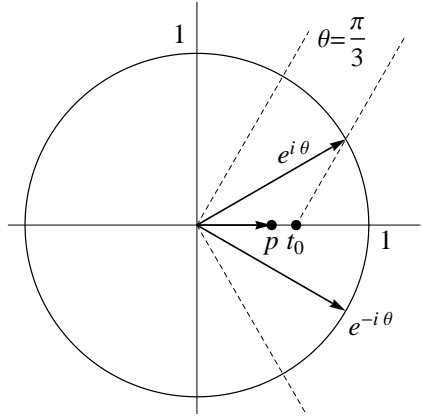}\,
\caption{The argument of $e^{i\theta}-p$ is less than that of $e^{i\theta}-t_0$ for $0<\theta<\pi/3$}
\end{figure}
Now, it is clear that
$|e^{i\theta}-p|<1$, and thus the condition~\eqref{p2} is automatically satisfied.
We alse see that  the argument $\theta^\prime$ of $e^{i\theta}-p$ satisfies the condition $|\theta^\prime |<\frac {\pi}3$. Then we have that
\[
\frac {1+b-p+c-p}{|e^{i\theta}-p|}=\frac {p_{\theta}-2p}{|e^{i\theta}-p|}
=\frac {(e^{i\theta}-p)+(e^{-i\theta}-p)}{|e^{i\theta}-p|}=\frac{|e^{i\theta}-p|(e^{i\theta^\prime}+e^{-i\theta^\prime})}{|e^{i\theta}-p|}
=p_{\theta^\prime}.
\]
Therefore, the condition~\eqref{p1} is also satisfied, and  interior points of $\Phi[a,b,c;\theta]$ are not co-optimal.
The other cases for $\theta$ are similar. We summarize the results in TABLE 1.

\begin{table}[h!]
\begin{tabular}{ccccccccccc}
  \hline\hline
 & & &\multicolumn{3}{c}{(Co-)Spanning property} & & &\multicolumn{3}{c}{(Co-)Optimality}\\\cline{4-6}\cline{9-11}
Faces & &  &Span. &   Co-span. &Bi-span.& & &Opt. &Co-opt. &Bi-opt.\\\hline
$ f_{\rm abc}^{\theta},f_{\rm ab}^{\theta}, f_{\rm ac}^{\theta}, f_{\rm bc}^{\theta}, e_{\rm a}^{\theta}, e_{\rm b}^{\theta},e_{\rm c}^{\theta}$
& & &N&N&N& &  &N&N&N\\
 $e_{\rm ab}^{\theta},e_{\rm ac}^{\theta},v_{(p_{\theta},0,0)}$
& & &N&Y&N& & &N&Y&N\\
 $e_t^{\theta}, v_{(0,t,1/t)}$
& & &Y&N&N& & &Y&N&N\\
$v_{(1,0,p_{\theta}-1)},v_{(1,p_{\theta}-1,0)}$
& & &N&Y&N& & &Y&Y&Y\\
$v_{(a(t),b(t),c(t))}$
& & &Y&Y&Y& & &Y&Y&Y\\
 \hline\hline
\end{tabular}\caption{Summary of (co-)optimality and (co-)spanning property for faces of the convex body $\Gamma^{\theta}$}
\end{table}

We note that the the bi-optimality automatically implies
indecomposability, and so we see that $v_{(1,0,p_\theta-1)}$,
$v_{(1,p_\theta-1,0)}$ and
$v_{(a_\theta(t),b_\theta(t),c_\theta(t))}$ give rise to an
indecomposable maps. This can be seen directly. We note that
$$
\lan W[p_{\pi-\theta}, t,{\textstyle\frac 1t};\pi-\theta], \Phi[a,b,c;\theta]\ran
=3( ap_{\pi-\theta}+bt+\frac ct -2).
$$
Assume $bc=(1-a)^2$ and take $t=\sqrt{\frac cb}$, to get
$$
\lan W[p_{\pi-\theta}, t,\textstyle\frac 1t;\pi-\theta], \Phi[a,b,c;\theta]\ran
=3(ap_{\pi-\theta}+2\sqrt {bc}-2)=3a(p_{\pi-\theta}-2).
$$
We note that $p_{\pi-\theta}<2$ if and only if $\theta\neq \pm\frac\pi 3,\pm\pi$.
Therefore, we see that $\Phi[a,b,c;\theta]$ is an indecomposable positive map
whenever the condition
$$
0<a\le 1,\quad a+b+c\ge p_\theta,\quad bc=(1-a)^2,\quad \theta\neq \pm\frac\pi 3,\pm\pi
$$
holds. Recall that we have already seen that positivity of $\Phi[a,b,c;\theta]$ implies decomposability
in the case of $\theta= \pm\frac\pi 3,\pm\pi$ for which $p_\theta=1$.

\section{Detecting PPT edge states}

We proceed to find an optimal entanglement witness which detects the
PPT entangled edge state \cite{kye_osaka} $W[p_\theta,b,\frac 1b,
\theta]$ for $-\frac \pi 3<\theta<\frac\pi 3$ with $\theta\neq 0$
and $b>0$. We note that if we put $z=(1,0,0\,;\, 0,1,0\,;\,
0,0,1)^\ttt$ and
$$
\begin{aligned}
w_1&=(0,\sqrt b,0\,;\, \textstyle\frac 1{\sqrt b}e^{i\theta},0,0\,;0,0,0)^\ttt,\\
w_2&=(0,0,0\,;\, 0,0,\sqrt b\,;0,\textstyle\frac 1{\sqrt b}e^{i\theta},0)^\ttt,\\
w_3&=(0,0,\textstyle\frac 1{\sqrt b}e^{i\theta}\,;\, 0,0,0\,;\,\sqrt b,0,0)^\ttt,\\
\end{aligned}
$$
then we see that
$$
\lan zz^*,W(p_\theta,b,\textstyle\frac 1b;\theta]\ran=
\lan w_iw_i^*,W(p_\theta,b,\textstyle\frac 1b;\theta]^\Gamma\ran=0.
$$
Therefore, the most natural candidate is
$$
W=\left(
\begin{array}{ccccccccccc}
1-\alpha    &\cdot   &\cdot  &\cdot  &1+e^{-i\theta}     &\cdot   &\cdot   &\cdot  &1+e^{i\theta}     \\
\cdot   &b-\beta &\cdot    &\cdot    &\cdot   &\cdot &\cdot &\cdot     &\cdot   \\
\cdot  &\cdot    &\frac 1b-\gamma &\cdot &\cdot  &\cdot    &\cdot    &\cdot &\cdot  \\
\cdot  &\cdot    &\cdot &\frac 1b-\gamma &\cdot  &\cdot    &\cdot    &\cdot &\cdot  \\
1+e^{i\theta}     &\cdot   &\cdot  &\cdot  &1-\alpha     &\cdot   &\cdot   &\cdot  &1+e^{-i\theta}     \\
\cdot   &\cdot &\cdot    &\cdot    &\cdot   &b-\beta &\cdot &\cdot    &\cdot   \\
\cdot   &\cdot &\cdot    &\cdot    &\cdot   &\cdot &b-\beta &\cdot    &\cdot   \\
\cdot  &\cdot    &\cdot &\cdot &\cdot  &\cdot    &\cdot    &\frac 1b-\gamma &\cdot  \\
1+e^{-i\theta}     &\cdot   &\cdot  &\cdot  &1+e^{i\theta}     &\cdot   &\cdot   &\cdot  &1-\alpha
\end{array}
\right),
$$
which equal to
$$
2\cos\frac\theta 2 W\left[\dfrac{1-\alpha}{2\cos\frac \theta 2},\
\dfrac{\frac 1b-\gamma}{2\cos\frac \theta 2},\
\dfrac{b-\beta}{2\cos\frac \theta 2},\
\pi-\frac \theta 2\right]
$$
since $1+e^{-i\theta}=2\cos\frac\theta 2 e^{-\frac{i\theta}2}$.

To begin with, we note that
$$
\begin{aligned}
\frac 13\langle W[p_\theta,\frac 1b, b,\theta], W\rangle
&=p_\theta(1-\alpha)+\frac 1b (b-\beta)+b (\frac 1b-\gamma)-p_\theta-2\\
&=-p_\theta\alpha-\frac\beta b-b\gamma<0.
\end{aligned}
$$
We search for $\alpha,\beta,\gamma>0$ so that $W$ is bi-optimal.
To do this, we look for
$\alpha,\beta$ and $\gamma$ satisfying the conditions
\[
\begin{aligned}
&2\cos\frac{\theta}2 (2-p_{\pi-\frac{\theta}2})<1-\alpha<2\cos\frac{\theta}2,\quad b-\beta>0,\quad \frac 1b-\gamma>0,\\
&(1-\alpha)+(b-\beta)+(\frac1b-\gamma)=2p_{\pi-\frac\theta 2}\cos\frac \theta 2,\\
&(b-\beta)(\frac 1b-\gamma)=\left(2\cos\frac\theta 2-(1-\alpha)\right)^2.
\end{aligned}
\]
from Theorem~\ref{thm:spanning}.
For the simplicity, we put
$$
t:=\cos\frac{\theta}2,\quad
\tilde \alpha=1-\alpha,\quad
\tilde \beta=b-\beta,\quad
\tilde \gamma=\frac 1 b-\gamma.
$$
Then we have $\frac {\sqrt3}2<t<1$ and $p_{\pi-\frac{\theta}2}=t+\sqrt{3(1-t^2)}$.
Now, the problem is reduced to look for $\tilde \alpha,\tilde \beta$ and $\tilde \gamma$ satisfying the conditions
\begin{align}
\label{cond:oew1}&2t(2-t-\sqrt{3(1-t^2)})<\tilde \alpha <2t,\\
\label{cond:oew2}&\tilde \beta+\tilde \gamma = 2t(t+\sqrt{3(1-t^2)})-\tilde\alpha,\quad
\tilde \beta \tilde \gamma=(2t-\tilde\alpha)^2,\quad
\tilde \beta>0,\quad \tilde \gamma>0.
\end{align}
It is easy to see that
\[
1<t+\sqrt{3(1-t^2)}<\sqrt 3,
\]
for $\sqrt 3/2<t<1$.
Therefore, if we choose $\tilde \alpha$ satisfying the condition~\eqref{cond:oew1} for each $\frac {\sqrt3}2<t<1$, then we see that
\[
\tilde \beta +\tilde \gamma >0\ \text{ and }\  \tilde \beta \tilde \gamma>0
\]
 in \eqref{cond:oew2} and
\[
(\tilde \beta+\tilde \gamma)^2-4\tilde \beta\tilde \gamma =
-[\tilde \alpha -2t(2-t-\sqrt{3(1-t^2)})][3\tilde \alpha-2t(2+t+\sqrt{3(1-t^2)})]\ge 0.
\]
Consequently, we can find $\tilde \beta$ and $\tilde \gamma$ as positve roots of the quadratic equation
\begin{equation}\label{cond:oew3}
x^2 -\left [2t(t+\sqrt{3(1-t^2)})-\tilde \alpha\right] x+(2t-\tilde \alpha)^2=0.
\end{equation}
This completes the proof of the following:

\begin{theorem}
For each $0<|\theta|<\frac{\pi}3$ and $b>0$,
let $\tilde \alpha$ be a positve number satisfying the condition~\eqref{cond:oew1} with $t=\cos\frac{\theta}2$,
and $\tilde \beta$ and $\tilde \gamma$ be roots of the quadratic equation~\eqref{cond:oew3}. Then
\[
\frac{2\cos(\theta/2)}{3(\tilde \alpha+\tilde \beta+\tilde \gamma)}W\left[\frac{\tilde \alpha}{2\cos(\theta/2)},
\frac{\tilde \beta}{2\cos(\theta/2)},\frac{\tilde \gamma}{2\cos(\theta/2)},\pi-\frac{\theta}2\right]
\]
is an optimal PPTES witness which detects $W[p_{\theta},b,\frac1 b,\theta]$.
\end{theorem}

We note that a general method had been suggested in \cite{lew00} to construct entanglement witnesses detecting
a given entangled state. If we follow this method for
$W[p_\theta,b,\frac 1b, \theta]$, then we get the above $W$ with $\alpha=\beta=\gamma$.
But, it turns out that this method does not give us an {\sl optimal} PPTES witness in general.
In fact, one can show that if $W$ is an optimal PPTES witness with positive $\alpha=\beta=\gamma$ then
we have the restriction
$$
b+\frac 1b \le 2-\sqrt{3}+\sqrt{6\sqrt{3}-6}
\quad {\text{\rm and}}\quad
\cos\frac{\theta}2\le\frac 18 (3+\sqrt{21}).
$$
That is why  we consider the above $W$ with different $\alpha,\beta$ and $\gamma$ for the full generality.

\section{Conclusion}

We determined positivity of linear maps with four parameters which are of Choi map types involving
complex entries. We also determined
their optimality, co-optimality, spanning property and co-spanning property. In this way, we
found parameterized examples of indecomposable positive linear maps with the bi-spanning properties.
They are optimal PPTES witnesses, which are \lq nd-OWE\rq s in the sense of \cite{lew00}.
Optimality is not so easy to determine for a given positive linear map, because we do not know the whole facial structures
of the convex cone $\mathbb P_1$ consisting of all positive maps. The spanning property is stronger than optimality and relatively easy to
check. We suggest a general method to check optimality of a positive map $\phi$:
We first find all extremal completely positive maps in the smallest exposed face
of $\mathbb P_1$ containing $\phi$, and check if they belong to the smallest face containing $\phi$.

The optimal PPTES witnesses we constructed
detect two qutrit PPT entangled edge states of type $(6,8)$ in \cite{kye_osaka},
whose existence had been a long standing question \cite{sbl}.
We report here one interesting byproduct of our construction. Our constructions give counter-examples to the
conjecture \cite{korbicz} regarding the structural physical approximations, which claims
that the SPA of an optimal entanglement witness is separable.
Several authors
\cite{aug_bae,chru_pyt,qi}
checked recently various kinds of entanglement witnesses to support the conjecture.
In the forth-coming paper \cite{ha_kye_spa}, the authors  will consider the SPA conjecture
in a systematic way. We introduce the notions of positive type and copositive type
for entanglement witnesses depending on the distances to the positive part and the copositive part.
We will show that if the SPA of an entanglement witness is separable then it must be of copositive type,
and so the SPA is meagingful only for those of copositive types.
Our construction in this paper shows that the SPA conjecture does not hold even in cases of copositive types.

\color{black}

\end{document}